\documentstyle[twocolumn,aps,prl,epsf]{revtex}
\newcommand{\beq}{\begin{equation}}
\newcommand{\eeq}{\end{equation}}
\newcommand{\beqa}{\begin{eqnarray}}
\newcommand{\eeqa}{\end{eqnarray}}

\begin{document}

\def\rl{\rangle \langle}
\def\openone{\leavevmode\hbox{\small1\kern-3.8pt\normalsize1}}
\def\RR{{\rm I\kern-.2emR}}
\def\tr{{\rm tr}}
\def\ce{{\cal E}}
\def\cc{{\cal C}}
\def\ci{{\cal I}}
\def\cd{{\cal D}}
\def\cb{{\cal B}}
\def\cn{{\cal N}}
\def\ct{{\cal T}}
\def\cf{{\cal F}}
\def\ca{{\cal A}}
\def\cg{{\cal G}}
\def\cv{{\cal V}}
\def\cc{{\cal C}}
\def\rhon{\rho^{\otimes n}}
\def\on{^{\otimes n}}
\def\pn{^{(n)}}
\def\pnp{^{(n)'}}
\def\tcd{\tilde{\cal D}}
\def\tcn{\tilde{\cal N}}
\def\tct{\tilde{\cal T}}
\def\id{\frac{I}{d}}
\def\pthang{\frac{P}{\sqrt{d \tr P^2}}}
\def\psirq{\psi^{RQ}}
\def\rhorq{\rho^{RQ}}
\def\rhorqp{\rho^{RQ'}}
\def\ra{\rangle}
\def\la{\langle}
\def\QED{\mbox{\rule[0pt]{1.5ex}{1.5ex}}}
\def\proof{\noindent\hspace{2em}{\it Proof: }}
\def\endproof{\hspace*{\fill}~\QED\par\endtrivlist\unskip}
\def\nn{\nonumber}

\newcommand{\half}{\mbox{$\textstyle \frac{1}{2}$} }
\newcommand{\ket}[1]{| #1 \rangle}
\newcommand{\bra}[1]{\langle #1 |}
\newcommand{\proj}[1]{\ket{#1}\! \bra{#1}}
\newcommand{\outerp}[2]{\ket{#1}\! \bra{#2}}
\newcommand{\inner}[2]{ \langle #1 | #2 \rangle}
\newcommand{\melement}[2]{ \langle #1 | #2 | #1 \rangle}
\newcommand{\matel}[2]{ \langle #1 | #2 | #1 \rangle}
\newcommand{\expect}[1]{\langle #1 \rangle}

\date{\today}
\title{Quantum secure identification using entanglement and catalysis}
\author{Howard Barnum}
\address{
School of Natural Science and 
Hampshire College,
Amherst, MA 01002, USA \\
email: {\tt hbarnum@hampshire.edu}
}
\maketitle


%

\begin{abstract}
I consider the use of entanglement between two parties 
to enable one to authenticate her identity to another over
a quantum communication channel.  Exploiting the phenomenon of 
entanglement-catalyzed transformations between pure states
gives a potentially reusable entangled identification token.
In analyzing this, I consider
the independently interesting problem of 
the best possible approximation to a given pure entangled
state realizable using local actions and classical communication by
parties sharing a different entangled state.
\end{abstract}
\vspace{.1 in}
\vspace{.2 in} \narrowtext

\section{Introduction}
The protocols to be presented in this paper give Alice
and Bob a way of identifying (``authenticating'')
each other, using quantum and classical
channels.  
The protocols are such that if Alice and Bob can successfully complete one,
Alice is convinced that Bob (or someone who has stolen his identification 
token) is on the other end of the quantum communication
channel.  The classical analogue of this can be done by having Bob reveal,
over a classical channel, a secret which Alice and Bob had previously 
securely shared.   The quantum protocols use shared entangled states
as the counterpart of shared secret key.
The quantum version of the protocol differs from the
classical protocol in potentially useful ways.  At the most elementary 
level, it authenticates the presence of one {\em quantum} laboratory with
certain causal relations to another on a network.  (This is accomplished 
using quantum information exchange;  lest it be thought impossible, we note
that this might also be
accomplished via classical information exchange, using quantum nonlocality.)
There is protection,
via the no-cloning theorem, against copying of the authentication token. 
Similar ideas may provide theft-detection capability, much as standard
quantum cryptography provides security against eavesdropping.  Most 
interestingly, the protocol based on using entangled ``catalyst'' states
may provide reusable authentication tokens, although the security properties
when the tokens are reused will require careful analysis.

\section{The task}
The authentication task considered in this paper is widely
performed in practice, with computer login sequences and automatic teller
PIN protocols among the most common examples.
It arises in settings where
the channel is considered sufficiently
tamper-proof by the users, but they wish to defend against a certain kind
of ``terminal'' insecurity:  the possibility that the workstation
(or quantum lab) has been taken over by an impostor.  For this to be
relevant to the authenticity of a session that follows authentication, 
there must be some assumption that control of the terminal has a high
degree of intertemporal correlation.  Perhaps there is some probability
per unit time that the control of the terminal will be seized by an 
impostor.  The probability of authorized control of the terminal then
decays exponentially with time, but successful login-style authentication at a
later time resets the the probability to one, which could provide (with a 
low enough decay rate) sufficiently
increased security for a session bounded in time
and starting with the authentication.  Or, it may be assumed that once
Alice is logged in, transmission remains authentic for the entire session;
violations of 
terminal security are assumed to require stealth on the part of the 
impostor, and be infeasible (or low-probability) 
until Alice has left the terminal.

Such protocols are sometimes used as a prelude to 
further communication, for example in a login sequence for remote use of
a computer.  Without assumptions such as those of the previous paragraph,
these protocols do not provide a solid guarantee of
authenticity for the rest of the session.  Someone could allow you to 
log in, then block your access and tap the line for their own purposes.
Demonstrating authenticity for the entire session
in the presence of such a threat requires methods such as the protocol
based on universal hash functions \cite{Wegman81a,Carter79a}.
(Such protocols are used, for example, to authenticate the 
{\em classical} communication 
involved in quantum key distribution protocols.)
In these protocols the  degree of security  
associated with shared secret key bits may be transferred to the authenticity
of the message, by hashing the message and the secret key using a certain
type of universal hash function, and sending the hash along with the
(unencrypted) message over the public channel.  The receiver 
then hashes the message
with his copy of the secret key, and compares to the sent hash;  tampering
(with either message or hash or both) by someone without knowledge of the key
is overwhelmingly likely to yield a tampered-with hash 
which is not the valid hash of the authentic message.  

Whether the
authenticity
of the quantum
authentication tokens discussed in this paper, 
which are a counterpart of the classical notion of 
shared secret key, can be transferred to an entire communication session
in the manner of the universal hashing protocol described above, is an 
extremely interesting question.  A simple adaptation of the 
classical protocol would
seem difficult in the case in which the user who logs in may wish to act
as a conduit for quantum states unknown to her, since the hashing protocol
requires two copies of the message, one to be hashed and sent and 
the other to be sent unaltered.  The preparation-visible case seems more
promising, although this is a severe restriction.

\section{Quantum authentication using entanglement}

Rather than using a shared classical secret, Alice and Bob
authenticate via a shared entangled quantum state.
Unlike the key in a classical scheme, this quantum key need not be secret, 
although keeping it secret might make it harder to steal (particularly if it 
is not maximally entangled), and might have other advantages in a many-user
setup.
Perhaps the simplest scheme is for Alice and Bob to share some maximally
entangled states;  Alice, at least, knows which states.  
To authenticate, Bob sends Alice his half of some of 
the states;  Alice then performs tests on them to assure herself they
are indeed the specified Bell states.  Say they are all the same Bell state;
Bob sends $N$ of them, and Alice measures them in the Bell basis;  by
increasing $N$, she can distinguish these states more and more reliably 
(at a rate which depends on the noise in the overall process)
from any states that could come from an 
impostor which are actually not entangled with her states.  
In fact, in a $d$-dimensional Hilbert space the maximum 
matrix element $\matel{\psi}{\rho}$ for a separable $\rho$ 
to pass a test for being a maximally entangled
$|\psi\rangle$ is $1/d$ (this occurs where $\rho$ is pure, and is 
any one of the product states occuring in the Schmidt decomposition).
Thus security increases exponentially in the number of qubits
($\log_2{d}$) used.


The quantum scheme differs from its classical counterpart in several
potentially useful respects.
With a classical secret, an impostor
would have to break into Bob's classical storage area and copy his secret
key; it is physically possible to do so without evidence of the break-in.
The impostor and Bob could both communicate with Alice, who might not
realize anything was wrong for quite a while.  Both would be able to 
authenticate.  By contrast, an impostor who stole Bob's half of a 
quantum authentication key would have to be able to always divert
the authentication portion of a communication session 
to himself, should he wish to
allow Alice and Bob to communicate unaware of the loss of their
authentication
token.
The quantum scheme is also more directed:
steal Alice's  classical (symmetric-crypto) key and you 
have what 
Alice and Bob share;  steal Alice's half of an entangled state and you 
still can't pretend to Alice that you're Bob.
The no-cloning theorem not only makes undetected theft of key more
difficult, but also protects a stolen key from dissemination to wide
sectors of the underworld.  Steal Alice's classical key, 
and you can distribute it to 
your henchmen all over, who can pretend to be Alice when and where they
want.  Steal her half of an entangled state, and there's no way of 
distributing it among your henchpeople (or henchthings, for that matter).

Either the maximally-entangled-state quantum authentication protocol
or the classical shared-secret protocol uses up some of the shared key
each time authentication takes place.  By contrast, a catalysis protocol
does not.  The fact that the shared key is (or can be)
used up each time authentication takes place renders the EPR protocol
better for certain purposes:  for example, for banknotes it could be 
good that the state
is returned to the bank and destroyed in the authentication process.
Such entangled ``money'' would have the desirable properties 
of transferability
and uncopyability, but not the properties of untraceability or
anonymity, nor does the 
present discussion provide a protocol for verification of its value, let
alone nondestructive verification of its value, by 
third parties; it is far from constituting proper quantum cash.  Schemes
more suitable for quantum cash were proposed by Wiesner \cite{Wiesner83a}
and Bennett, Brassard, Breidbart and Wiesner \cite{Bennett82a};
they involve nonorthogonality rather than entanglement.
For other purposes, however, it may be good to have
a reusable i.d. token.  The catalysis protocol presented in Section
\ref{sec: catalysis protocol}
may provide such a token.  The next section provides background for
the understanding of catalysis.

\section{LOCC-convertibility and catalysis}
Nielsen \cite{Nielsen98b} showed that 
a pure state
$\ket{\psi}$ in a Hilbert space $A \otimes B$ is convertible 
to another, $\ket{\chi}$, by local
actions and classical communication (LOCC), if and only if the
nonincreasingly-ordered
eigenvalues of 
the reduced density matrix of $\ket{\chi}$ majorize those of $\ket{\psi}$.
(These are sometimes called the OSC's, ordered Schmidt coefficients,
of the pure states.  I will use the notation 
$\lambda_i(\proj{\eta})$ for the ordered eigenvalues
of the {\em reduced} density matrix of $\proj{\eta}$, rather than for
those of the density matrix of $\proj{\eta}$ itself as in more
standard notation.)
That is, the target state's reduced 
density matrix is ``less mixed'';  it is natural
to suppose this means it is less entangled, and the theorem confirms this
intuition.
I will write this relation (which is a partial ordering on the pure states)
$\ket{\chi} \preceq \ket{\psi},$ and write $\ket{\psi} \bowtie \ket{\chi}$
when neither $\ket{\psi} \preceq \ket{\chi}$ nor $\ket{\chi} \preceq
\ket{\psi}.$   (The relation ``$\preceq$'' 
may thus be read ``is less entangled than.'')
I will also use the obvious ordering  this induces on 
one-dimensional projectors on these states, and extend it to density
matrices in the manner of Vidal \cite{Vidal99a}.  
He defined $d-1$ entanglement measures 
$E_k$, which are sums of the 
{\em k lowest} Schmidt eigenvalues, and are nonincreasing under LOCC.
Define $S_k(\proj{\eta}) := \sum_{i=1}^k \lambda_i(\proj{\eta})\;.$
Also define $E_{k-d+1}(\proj{\eta}) := 1 - S_k (\proj{\eta}) 
\equiv \sum_{k-d+1}^d \lambda_i(\proj{\eta})\;$.
Extend this to mixed states via:
\beq
S_k(\rho) := \max_{\{t_j, |\eta_j\rangle |\sum_{j} t_j \proj{\eta_j} 
= \rho\} }
\sum_j t_j S_k(\proj{\eta_j})\;.
\eeq
Equivalently, one could define
\beq
E_k(\rho) := \min_{\{t_j, |\eta_j\rangle |\sum_{j} t_j \proj{\eta_j} 
= \rho\} }
\sum_j t_j E_k(\proj{\eta_j})\;.
\eeq
Define
\beq \label{generalized majorization}
\rho_1 \succeq \rho_2 := \forall k, E_k(\rho_1) \ge E_k(\rho_2)\;.
\eeq
This is the extension of the majorization-based
``more entangled than'' relation to mixed states.
Vidal \cite{Vidal99a} showed that the $E_{k-d+1} := 1 - S_k$
are entanglement montones, 
hence cannot be increased by LOCC.
Jonathan and Plenio
\cite{Jonathan99a}
state (indeed, they make a somewhat more general statement; cf. also
\cite{Hardy99a})
that the partial ordering 
(\ref{generalized majorization}) in fact coincides with LOCC-convertability.
This is, of course, a generalization of Nielsen's theorem to mixed states.
Jonathan and Plenio \cite{Jonathan99b} have used Nielsen's Theorem
to show the existence
of pairs of states $\ket{\psi} \bowtie \ket{\chi}$, such that if Alice and
Bob share a particular entangled state $\ket{\phi}$, they may nonetheless
convert $\ket{\psi}$ into $\ket{\chi}$ by LOCC, while retaining $\ket{\phi}$
unchanged at the end of the process.  In this paper, this phenomenon is
exploited to give a potentially reusable quantum identification token.

\section{The catalysis protocol} \label{sec: catalysis protocol}

Here, Alice and Bob share a catalyst state $\ket{\phi}$.  There are
incommensurate states $|\phi_1\rangle$ and $|\phi_2\rangle$ such that
in the presence of the catalyst,  
$|\phi_1\rangle$ can be converted to $|\phi_2\rangle$, while
retaining $|\phi\rangle$.  For Bob to authenticate himself to 
Alice, Alice makes $|\phi_1\rangle$ in her laboratory, and sends 
half of it to Bob.  They then go through the steps, involving 
local measurements, one-way communication of measurement results, 
and local operations conditional on those measurement results,
which convert $|\phi_1\rangle$ to $|\phi_2\rangle$.  If Alice's quantum
channel leads to a cheating Derek,
who does not possess any subsystem involved in the catalyst state,
by Nielsen's result this conversion cannot succeed.  So the protocol
continues by having Bob return to Alice the B half of the 
system which they
were to convert into $|\phi_2\rangle$.  She then measures the projector
onto $|\phi_2\rangle$.  If (with the idealization
of perfect operations and measurements) she ever gets the result ``0'',
she knows that Bob was not involved in the protocol (assuming Bob would
always implement the protocol correctly were he involved).
Usually, even if an impostor is involved, the measurement result will be
``1''.  But the small probability of ``0'' in that case can be amplified
by repeating the protocol.  This has no cost in the stored catalyst state,
though it has a cost (polynomial in the desired accuracy)
in quantum communication.

A weak upper bound on the one-shot probability of error (of getting ``1'' 
from an impostor) can be gotten by considering 
the (convex) set of less entangled states (those whose OSC's majorize
$|\phi_1\rangle$'s).
Let $\rho^*$ be the closest of these states (in 
the $L_1$ norm distance which corresponds to error probability) to 
$|\phi_2\rangle$;  then $p_e = \melement{\phi_2}{\rho^*}$.   
This can be used to obtain, for a specified
$\epsilon$, an expression, polynomial in $1/\epsilon$,
for a number of repetitions guaranteed to give error probability below
$\epsilon.$  It is a bound because only the less entangled states
are accessible via a protocol between Alice and an 
impostor Derek unentangled with those of Alice's systems involved in 
the protocol.  It is a weak bound because choosing the nearest of those
states means  considering
Alice as conniving with Derek to fool herself.  In the actual situation, Alice
will not perform her part of a protocol for converting
$|\phi_1\rangle$ to $\ket{\phi^*}$, but rather will still perform her part
of the protocol for converting $|\phi_1\rangle$ to $\ket{\phi_2}$.
As Chris Fuchs pointed out to me, when Alice does the correct protocol
in the presence of the impostor, she will of course wind up with the
same reduced density matrix she would have if there had been no 
impostor; i.e., the reduced density matrix of $|\phi_2\rangle$.
So, we may now look at the closest state, not merely among the
less entangled states, but in the (still convex) set of
less entangled states 
having the same density matrix for Alice as $|\phi_2\rangle$.
Since $|\phi_2\rangle$ and local unitary transformations of it 
are not in the majorized set, the closest such state will be mixed.

Our weak bound on the error probability is given by 
\beq
\max_{\rho \preceq \proj{\phi_1}} 
\matel{\phi_2}{\rho}\;.
\eeq
This problem would be very much simplified if we could assume the
optimal state $\rho^*$ were pure,  but it is not obvious this should
be so.  It is clear that the majorization constraint on $\rho$ prevents
us from attaining $1$ in this maximization, but in order to use the
protocol in a particular instance, we need an upper bound below $1$.
One more tractable 
upper bound comes from considering the 
$d-1$ maximization problems in which only one of the
majorization constraints is imposed, and taking the lowest of these
maxima.  
\beqa
\min_{k=1,...,d-1} \max f(\rho) :=  \matel{\phi_2}{\rho} \nonumber \\
{\rm subject~~to~} \nonumber \\
E_k(\rho) \le \zeta_k\;.
\eeqa
Here $\zeta_k := E_k(\proj{\phi_1}).$
Because of the definition of $E_k$ as a minimum over ensembles,
we can recast the inner maximization as a maximization of $f$ over 
all ensembles of pure states:
\beqa \label{innermax}
\max f(s_1,...,s_n, |\eta_1\rangle, ..., |\eta_n\rangle)
:= \sum_j s_j |\inner{\phi_2}{\eta_j}|^2 \nonumber \\
{\rm subject~~ to} \nonumber \\
\sum_j s_j E_k(\proj{\eta_j}) \le \zeta_k\;.
\eeqa
Here $s_j$ are probabilities, and $|\eta_j\rangle$ pure states.
$n$ may vary, but there will be a bound from a Davies-type 
argument.
Moreover, there will be a maximum for which all the states 
$\ket{\eta_j}$ are Schmidt-codiagonal with $\ket{\phi_2}$.
This follows from two observations. First, 
Lemma $1$ below, which implies that each term in the objective 
function is maximized (for fixed $q_j$), by an $\ket{\eta_j}$
Schmidt-codiagonal with $\ket{\phi_2}$. 
Second, the fact that the local unitary transformation
required to get an arbitrary $|\eta_j\rangle$ into that form 
has no effect on 
the value of the constraint function, since $E_k$ 
is invariant under local unitaries.

\noindent
{\it Lemma 1:}

\noindent
For fixed Schmidt coefficients, 
the pure state $|\chi\rangle$ which maximizes 
$|\inner{\phi}{\chi}|^2$
has the same Schmidt basis as 
$|\phi\rangle$.  

\noindent 
{\it Proof:}

{\noindent}We 
show this by mapping the pure states of the
$d^2$-dimensional system $AB$ onto operators on a $d$-dimensional
Hilbert space, in such a way that the inner product in the tensor
product vector space becomes the Hilbert-Schmidt inner product of
the operators.  Denote such a map by $\sigma$, and use the notation
\beqa
\sigma(|\chi\rangle) := G_\chi.
\eeqa
Then 
\beqa
\inner{\phi}{\chi} \equiv \tr~G_\phi^\dagger G_\chi\;.
\eeqa
We may change to an arbitrary Schmidt basis by local unitaries.
In this formalism, and writing the reduced density matrices
as $\phi_1$ and $\chi$ respectively, unitaries corresponding 
to changing $|\chi\rangle$'s Schmidt bases in the $A$ and $B$
system are mapped, one-to-one, onto unitaries
$U$ and $W$ such that $G_\chi = U \chi^{1/2} W$.  Let 
$\chi$ and $\phi_1$ be diagonal in the same basis, for simplicity;
by varying over $U$ and $W$ we vary over $G$ operators corresponding
to all Schmidt bases, for given eigenvalues.  Thus 
\beqa
{\max}_{{\rm unitary} U,W} |\tr \phi_1^{1/2} U \chi^{1/2} W | 
\nonumber \\
= {\max}_{{\rm unitary} V,Y} |\bra{\phi_1}{V \otimes Y}\ket{\chi}|\;.
\eeqa    
A result of Von Neumann \cite{VonNeumann37a,Horn85a}
states that this maximum occurs where
$\phi_1^{1/2}$ and $\chi^{1/2}$ are codiagonal, and their eigenvalues
are matched in order of size.
This proves Lemma 1.

Now we 
show that there is an ensemble solving (\ref{innermax})
which contains just one pure state.
As just argued, we may confine our attention to ensembles of states
Schmidt-codiagonal with $\ket{\phi_2}$.  Thus, 
the squared inner products whose average is the objective function 
in (\ref{innermax}) are just squared Bhattacharyya overlaps 
\beq
B^2({\bf p},{\bf q}) := 
(\sum_i \sqrt{p_i q_i})^2
\eeq
between
the Schmidt eigenvalues $p_i$ of $|\phi_2\rangle$ and those ($q_i$)
of the states
$\ket{\eta_j}$;  since $B^2$ is concave in one argument
\cite{Fuchs97c},
replacing the $s_j$-ensemble by the vector with the $s_j$-averaged Schmidt
coefficients will increase the objective function.  Moreover, this will
keep the value of the constraint function unchanged.  To be explicit,
define 
\beq
\ket{\overline{\eta}} := \sum_i \sqrt{\sum_j s_j q_{ij}}
\ket{i}\ket{i}\;
\eeq
where $q_{ij}$ are the ordered Schmidt coefficients of $\ket{\eta_j}$.
Then 
\beqa
|\inner{\phi_2}{\overline{\eta}}|^2 = 
(\sum_i \sqrt{p_i \sum_j s_j q_{ij}})^2 \ge \sum_j s_j 
(\sum_i \sqrt{p_i q_{ij}})^2 
\nonumber \\
\equiv \sum_j s_j 
|\inner{\phi_2}{\eta_j}|^2\;.
\eeqa
while (for any $k$)
\beqa
S_k(\proj{\overline{\eta}}) = \sum_{i=1}^{k} (\sum_j s_j q_{ij})
= \sum_j s_j (\sum_{i=1}^{k} q_{ij})  \nonumber \\
= 
\sum_k s_j S_k(\proj{\eta_j})\;.
\eeqa

This argument doesn't immediately extend to the multi-constraint
problem.  Each $E_k$ is defined by an independent 
minimization over ensembles;
therefore one can't recast the multiconstraint 
problem as an single maximization
over ensembles.

If we nevertheless assume the solution is pure in the original 
multi-constraint problem, 
it reduces to:

\beqa
\max_{q_1,...,q_d} \sum_i \sqrt{p_i q_i}  \nonumber \\
{\rm subject~~to:} \nonumber \\
\zeta_k - \sum_{i=1}^k q_i \le 0~~(k=1,...,d),\nonumber \\
 -q_i \le 0\;, \nonumber \\
\sum_i q_i - 1 \le 0\;, 
\eeqa
where $p_i$, $q_i$, $r_i$ are the reduced density matrix eigenvalues
of $|\phi_2\rangle$, $|\chi\rangle$, and $|\phi_1\rangle$ respectively,
and $\zeta_k := \sum_{i=1}^k r_k$.
Since the objective function is concave in ${\bf q}$ and the feasible
set (since it is given by a conjunction of linear inequalities
$g_i({\bf q}) \le 0$))
is convex, the Kuhn-Tucker conditions are necessary and sufficient
for a maximum \cite{Varian78a} as long as constraint qualification holds.
These are:
\beq
\frac{\partial f({\bf q})}{\partial q_i} = \sum_i \lambda_i 
\frac{\partial g_i({\bf q}^*)}{\partial q_j}\;.
\eeq
Here 
$\lambda_i \ge 0$, with $\lambda_i = 0$ for slack constraints.

In our problem
the objective function $f({\textbf q}) := \sum_i \sqrt{p_i q_i}$
has derivatives
\beq
\partial f/ \partial q_i = (1/2) \sqrt{p_i/q_i}\;.
\eeq
The positivity constraints on the $q_i$ will be slack
because the partial derivatives go to $+ \infty$ at 
$q_i=0$.

Here is an example of the weak bound, using states from 
\cite{Jonathan99b}.  These are:
\beqa
|\phi_1\rangle = \sqrt{0.4} \ket{00} + \sqrt{0.4} \ket{11} + \sqrt{0.1}
\ket{22} + \sqrt{0.1}|33\rangle\; \nonumber \\
|\phi_2\rangle = \sqrt{0.5} \ket{00} + \sqrt{0.25} \ket{11} + \sqrt{0.25}
\ket{22}\;.
\eeqa
We will make the assumption that the optimal state is pure, which will
turn out to be correct in this particular case.
If we start with $|\phi_1\rangle$ and reallocate Schmidt coefficients from 
lower-weight to higher-weight basis states (so that the resulting state 
``majorizes'' $|\phi_1\rangle$, it is clear we should take all of the 
weight from $\ket{33}$ and move it to a higher-probability state, say
$\ket{22}$ to get its probability closer to $p_2$'s.
I chose to allocate it to $q_2$ because the derivative of the objective 
function was largest with respect to $q_2$, even after the reallocation of
all of $q_3$ to it.
No further increase in $q_2$ is possible, since the 
weight would have to come from $q_0$ or $q_1$, and that would violate
the majorization constraint $q_0 + q_1 = 0.8$.
Reallocating that $0.8$ optimally  
between the $q_0$ and $q_1$ (by equating derivatives
of the objective function with respect to them) yields 
$q_0 = \sqrt{8/15}, q_1 = \sqrt{4/15}$.  There is (of course)
still no advantage to reallocating $q_2$ to either of the
larger probabilities, since $\partial f/\partial q_2$ is still the largest
derivative.
Thus
\beq
|\chi^*\rangle = \sqrt{8/15}|00\rangle + \sqrt{4/15}|11\rangle
+ \sqrt{0.2}|22\rangle\;.
\eeq 
Hence $p_e = |\inner{\chi}{\phi_2}|^2 = .9964102....$
Moreover, here the only binding majorization constraint is 
$q_0 + q_1 \ge 0.8$.  Therefore, this also solves the problem with
that constraint alone, and so provides a genuine upper bound on
the protocol's error probability.
This is, of course, quite high, but since the error is one-sided it
can be exponentially suppressed by repetition.  
For example, after two thousand repetitions
we get $p_e = .0007522...$.  Still, computation of the stronger bound
obtained from consideration of Alice performing the actual protocol
is obviously desirable.   For practical purposes, it might be very
useful to find examples of catalyzable transformations 
$\ket{\phi_1} \rightarrow \ket{\phi_2}$ for which
the error probability of any local approximation of the transformation
is bounded much further below $1$.  It is natural to look in
higher-dimensional
systems for such examples.  Such pairs of states must be 
incommensurable, and Nielsen has conjectured that incommensurability
approaches being generic in high-dimensional systems.  Whether 
catalyzability is equally common (or whether it is measure-zero 
even for finite-dimensional systems) is an interesting question.
The distinguishability of a maximally entangled state from 
the separable states increases with dimensionality, with error-probability
approaching zero.  It is natural to ask 
whether the error probability for distinguishing 
states locally produced from a given state from some state 
locally producible
only with the aid of a catalyst
may be made to approach zero in some sequence of examples of increasing
dimension.

Whatever the ultimate judgement on the catalyzed protocol from the
cryptographic standpoint, the optimization problem discussed in this
section, of finding the closest approximation to $\ket{\phi_2}$ 
obtainable by local quantum operations and quantum communication,
given the state $|\phi_1\rangle$, is of independent interest.

\section{Security}
How secure is this protocol?  The error probability above gives a 
measure.  Since Alice may destroy her test states (the ones to be turned
from $|\phi_1\rangle$ into $\ket{\phi_2}$) after each use, coherent 
eavesdropping involving these states may seem unlikely to be of any
use in corrupting later uses of the protocol.  However, one must
investigate whether a Derek able to divert some of the quantum
communications,  and possibly also impersonate Alice or Bob 
during the classical 
discussion\footnote{Classical hashing techniques could be used 
to prevent this,
but this would require using up classical secret key, and, if crucial
to security, might obviate some advantages conferred by the quantumness
of the remainder of the protocol.},
could redirect the catalyst state
to himself and use it in later rounds.  If Derek is a true 
man-in-the-middle, he could receive Alice's 
test state, keep it, and send whatever
he wanted to Bob.  (He could even send half of a $\ket{\phi_1}$ state.)
If he could also impersonate Alice and Bob on the classical channel, 
then he could in fact do the protocol with Bob (who would be using
half of a Derek-supplied $\ket{\phi_1}$ state).  He could then send
the resulting $\ket{\phi_2}$ state on to Alice.  But that's not ``really''
a problem with the protocol:  Alice is indeed identifying herself.  
(Of course, Derek can jam her ability to i.d. herself, if he has this
setup, but men-in-the-middle can always do that.)  
But can Derek steal the catalyst?
It's somewhat plausible that he could.  For, he can send whatever state
he wants to Bob (even one entangled with a system Derek keeps), while keeping 
the state Alice sent him.  Bob will measure this other state jointly with
his end of the catalyst, and broadcast the result.  Bob will even send
the noncatalyst part of the measured system back to Derek.  
The basic elements present  (entanglement between Derek and
Bob, joint measurement by Bob of the system entangled with Derek and
Bob's part of the catalyst, and broadcast by Bob of the measurement
result) are also present in a protocol for teleporting Bob's part of
the catalyst state to Derek.  There is of course no guarantee (and
probably no particular reason to think) that this {\em is} such a
teleportation protocol.   Quite possibly, anything Derek does to 
steal the catalyst will be likely to make the conversion procedure
fail (after all, the simplest procedure for ensuring its success, 
which is for Derek to just act as a conduit for $\ket{\phi_1}$, 
cannot steal the catalyst).  But a full analysis is required before
claiming that the catalysis protocol is secure when reused 
(and reuse is necessary just to get the error probability acceptably
low, in the example given above).  It seems almost too good to be true
that a phenomenon such as catalysis, closely linked to properties (e.g. 
incommensurability, in the operational sense of non-interconvertibility
via LOCC) of 
finite-dimensional quantum states that disappear when the tasks defining
them are defined asymptotically on large numbers of copies,
should nevertheless be usable by repetition to achieve asymptotically
useful results in a cryptographic task.  But quantum information has
surprised us before.  And even if repeated
authentication via catalysis turns out
not to be secure, knowing this (and knowing why) 
will shed light on the phenonomenon of catalysis 
and related information-theoretic
concepts and tasks.

The teleportation attack scenario is not so worrisome in the
case in which only ``terminal security'' is at issue, which is the one
in which using the login protocol as a prelude to further communication
makes the most sense.  Nevertheless, it is worth investigating, if 
any level of channel insecurity exists:  it would
enable channel insecurity to become terminal insecurity over time.
Also, in a model with no channel insecurity, it is questionable
why one would want to use the catalysis protocol instead of the 
maximally-entangled protocol.  For then the problem 
of using up the supply of 
entangled key may be solved by sharing more entangled states over the
(supposedly secure) quantum channel during the period of high terminal security
initiated by authentication.  With any terminal insecurity, the 
security of the currently-used 
entangled states will decay with repeated sessions (some of the
new entangled states may occasionally be transmitted to an impostor).
Possibly this is not so for the catalyst states even with some 
terminal insecurity during a communication session, since once 
shared the catalyst states are never retransmitted;  here, again,
teleporting attacks are the issue.

\section{Conclusion}
I have given some protocols with which two parties may share an
entangled quantum state, and use it as a secure identification token.
The simplest protocols just involve sharing a maximally entangled state
of high-dimensional quantum systems;  authentication is accomplished by
one party's sending the other her half of the state, which can then be
distinguished by measurement from anything a disentangled impostor 
could present.  Advantages over classical shared secrets might include
theft detectability, uncopiability, and of course the ability to
authenticate a quantum laboratory's presence on a quantum network.
The possibility of a protocol based on the ability of 
some shared entangled states to catalyze certain transformations
between other shared entangled states (which would otherwise be 
impossible by local actions and classical communication) was also
introduced.  This occasioned some analysis of the problem, interesting
in itself, of the best LOCC approximation to such a transformation.
An interesting potential advantage of the catalysis protocol is 
repeatability without using up the identification token.  Showing
the security, or lack of it, of repeated use of this protocol could
illuminate several interesting areas of the theory of quantum information
and entanglement, in addition to 
shedding light on nature of the curious phenomenon of 
catalysis.

\begin{acknowledgments}
Supported in part by NSF grant 
\#PHY-9722614, and by a grant from the ISI Foundation, 
Turin, Italy, and Elsag-Bailey, a Finmeccanica company. 
Thanks to C. Bennett, C. Fuchs and L. Hardy
for discussions and encouragement.
\end{acknowledgments}





\end{document}